\documentclass[final,3p,times,twocolumn]{elsarticle}

\usepackage{amssymb}
\usepackage{amsmath}
\usepackage{float}
\usepackage{siunitx}
\usepackage{caption}
\usepackage{subcaption}
\usepackage[version=4]{mhchem}
\journal{Solid State Communications}

\begin{document}

  \begin{frontmatter}

    \title{Magnetic field induced polarization enhancement in the photoluminescence of MBE-grown \ce{WSe2} layers}

    \author[fuw]{Maksymilian Kuna}
    \author[fuw]{Mateusz Raczyński}
    \author[fuw]{Julia Kucharek}
    \author[TT]{Takashi Taniguchi}
    \author[KW]{Kenji Watanabe}
    \author[fuw]{Tomasz Kazimierczuk}
    \author[fuw]{Wojciech Pacuski}
    \author[fuw]{Piotr Kossacki}

    \affiliation[fuw]{organization={Institute of Experimental Physics, Faculty of Physics, University of Warsaw},
                addressline={ul. Pasteura 5}, 
                city={Warsaw},
                postcode={02-093}, 
                country={Poland}}

    \affiliation[TT]{organization={Research Center for Materials Nanoarchitectonics, National Institute for Materials Science},
                addressline={1-1 Namiki}, 
                city={Tsukuba},
                postcode={305-0044}, 
                country={Japan}}
                
    \affiliation[KW]{organization={Research Center for Electronic and Optical Materials, National Institute for Materials Science},
                addressline={1-1 Namiki}, 
                city={Tsukuba},
                postcode={305-0044}, 
                country={Japan}}

    \begin{abstract}
      We report an experimental study of the magnetic-field dependence of the optically pumped valley polarization in an epitaxial tungsten diselenide (\ce{WSe2}) monolayer grown by molecular-beam epitaxy (MBE) on a hexagonal boron nitride (hBN) substrate. Circularly polarized photoluminescence (PL) measurements reveal that applying a weak out-of-plane magnetic field, on the order of $0.1$ T, dramatically increases the effectiveness of the optical orientation of the emission associated with defect-bound localized excitons. We compare the obtained results with the earlier studies on the reference exfoliated monolayers, discussing both qualitative similarity as well as quantitative differences. Our observations are further supplemented by the results of time-resolved PL measurements, which confirm the pseudospin relaxation time of approximately 25 ps, a value significantly shorter than the $\approx$100 ps previously reported for mechanically exfoliated samples. 
    \end{abstract}

  \end{frontmatter}

\section{Introduction}
\label{introduction}
    The family of transition metal dichalcogenides (TMDs) represents a class of semiconductors with unique optoelectronic properties, particularly in their monolayer form. A crucial characteristic of these materials, such as \ce{WSe2}, is the transition from an indirect bandgap in bulk to a direct bandgap in a single atomic layer. This property, combined with a lack of inversion symmetry and a strong spin-orbit coupling, results in two distinct valleys ($K^+$ and $K^-$) in the first Brillouin zone, which can be selectively addressed using a circularly polarized light.
    
    A major challenge for practical valleytronic applications is the robustness of the valley information against various valley depolarization processes. Ideally, the system would retain the valley polarization from the optical excitation up to its readout in the form of the optical recombination of the exciton. However, experiments show that the circular polarization of the emitted light (which corresponds to the exciton valley) only weakly follows the polarization of the excitation, directly proving that, on the way, the valley degree of freedom undergoes some relaxation processes. Understanding of these valley relaxation processes has attracted a lot of research attention \cite{long-times, valleytronics-przeglad}, both theoretical and experimental. The focus of these studies was typically on the processes taking place in the relatively well-defined case of carriers already occupying the lowest energy bands, from which the recombination occurs.
    
    Much less attention was devoted to the relaxation processes occurring at early stages of the non-resonant excitation. This is arguably the more usual case, but poses significant challenge in rigorous description as when the photo-excited carriers have large excess energy, they can undergo a plethora of relaxation scenarios. For that reason, there is constant need for the experimental data which can shed more light on these relaxation processes.
    
    Previous studies \cite{smolen1} demonstrated that analyzing polarization-resolved photoluminescence of a \ce{WSe2} monolayer in a magnetic field under Faraday configuration offers unique insights into valley relaxation during post-excitation energy relaxation. A remarkable finding from this earlier work was that a weak out-of-plane magnetic field, on the order of 100 mT, could suppress intervalley scattering and thus enhance the degree of circular polarization of the emitted photoluminescence when exciting with a given circular polarization (FIPE -- field induced polarization enhancement). The properties of this effect led to the conclusion about the role of an intermediate dark state being an important stage of the relaxation cascade \cite{smolen1}. 
    
    So far, the FIPE effect was reported only for mechanically exfoliated \ce{WSe2} and \ce{WS2} monolayers \cite{smolen1, smolen2}. However, while exfoliated samples are considered to have high crystal quality, a pathway to scalable and uniform monolayers requires the use of modern fabrication techniques such as chemical vapour deposition (CVD) or molecular-beam epitaxy (MBE). In particular, the latter one has been shown to produce high-quality samples with exceptionally narrow exciton line and large-scale optical homogeneity, provided that the growth is done on a hBN covered substrate \cite{Pacuski}. 
    
    The fabrication technique itself, however, may affect the properties of the resulting monolayers. A notable differences between mechanically exfoliated samples and MBE-grown ones were found, e.g., with respect to the exciton recombination time \cite{kacper} or the valley relaxation seen by pump-probe experiments \cite{valley-dynamics}.
    
    Here, we investigate the effect of FIPE on MBE-grown samples in order to fully understand and control valley dynamics. We discuss the observed properties against the reference exfoliated monolayers reported in earlier works. Our data reveal that while the phenomenon persists, its underlying temporal dynamics is profoundly different. The observed quantitative differences in the magnetic field dependence point to a significantly faster pseudospin depolarization rate in the MBE-grown material, highlighting a direct link between the material's synthesis method and its valleytronic properties.

\section{Sample and experimental setup}
\label{sample}
    The \ce{WSe2} monolayer sample used in this study was grown by molecular-beam epitaxy (MBE) on hBN flakes, which were pre-exfoliated onto a \ce{SiO2}/Si substrate. This method, consistent with the procedure described in previous literature \cite{smolen1}, is known to produce samples with narrow excitonic lines and excellent optical homogeneity, setting them apart from materials grown on more disordered substrates that do not exhibit as perfect flatness as the h-BN\cite{MBE-growth}. Moreover, it's 2D structure ensures no dangling bonds are present, significantly enhancing the tungsten atoms mobility.
    
    The experiments reported here were conducted at cryogenic temperatures down to 5 K in helium exchange gas. The sample was subjected to magnetic field in Faraday geometry up to 10 T, however the crucial results are reported for the range below 0.5 T. Optical access to the sample was provided by means of free-beam optics, with a high-NA aspheric lens to focus the excitation laser down to a spot of 0.6-0.8 $\mu$m in diameter.
      
    The FIPE effect relies on measurement of the photoluminescence (PL). In our investigation, we performed both continuous-wave (CW) and time-resolved PL measurements. In both cases the combination of linear polarizers and half- and quarter-wave waveplates in both excitation and detection path allowed us to independently control the helicity of the exciting laser and the detected PL signal. The optical setup included optical filters (short-pass in the excitation path and long-pass in the detection path), which was used to ensure no direct contribution of the scattered laser light in the measured signal.
      
    For steady-state PL measurements, the sample was excited with continuous-wave lasers at wavelengths of 647 nm (1916 meV) and 680 nm (1823 meV). The signal was resolved by a 50-cm Czerny-Turner spectrometer and detected using a CCD camera.
      
    For time-resolved measurements, the sample was excited with a pulsed Ti:Al$_2$O$_3$ laser operating at 700 nm (1771 meV) with a repetition rate of 76 MHz and a pulse duration of approximately 150 fs. The temporal evolution of the PL signal was recorded using a streak camera equipped with a 30-cm spectrometer for spectral resolution.

\section{Results and discussion}
    The CW excited PL spectra of the monolayer, excited at different wavelengths, are presented in a Figure 1. The neutral exciton (X) dominates the spectra under non-resonant excitation, while the broad band of localized excitons (LE) is significantly enhanced under quasi-resonant pulsed excitation at 700 nm, which is closer to the LE emission energy.
    
    \begin{figure*}[htb]
        \centering
        \includegraphics[width=0.95\linewidth]{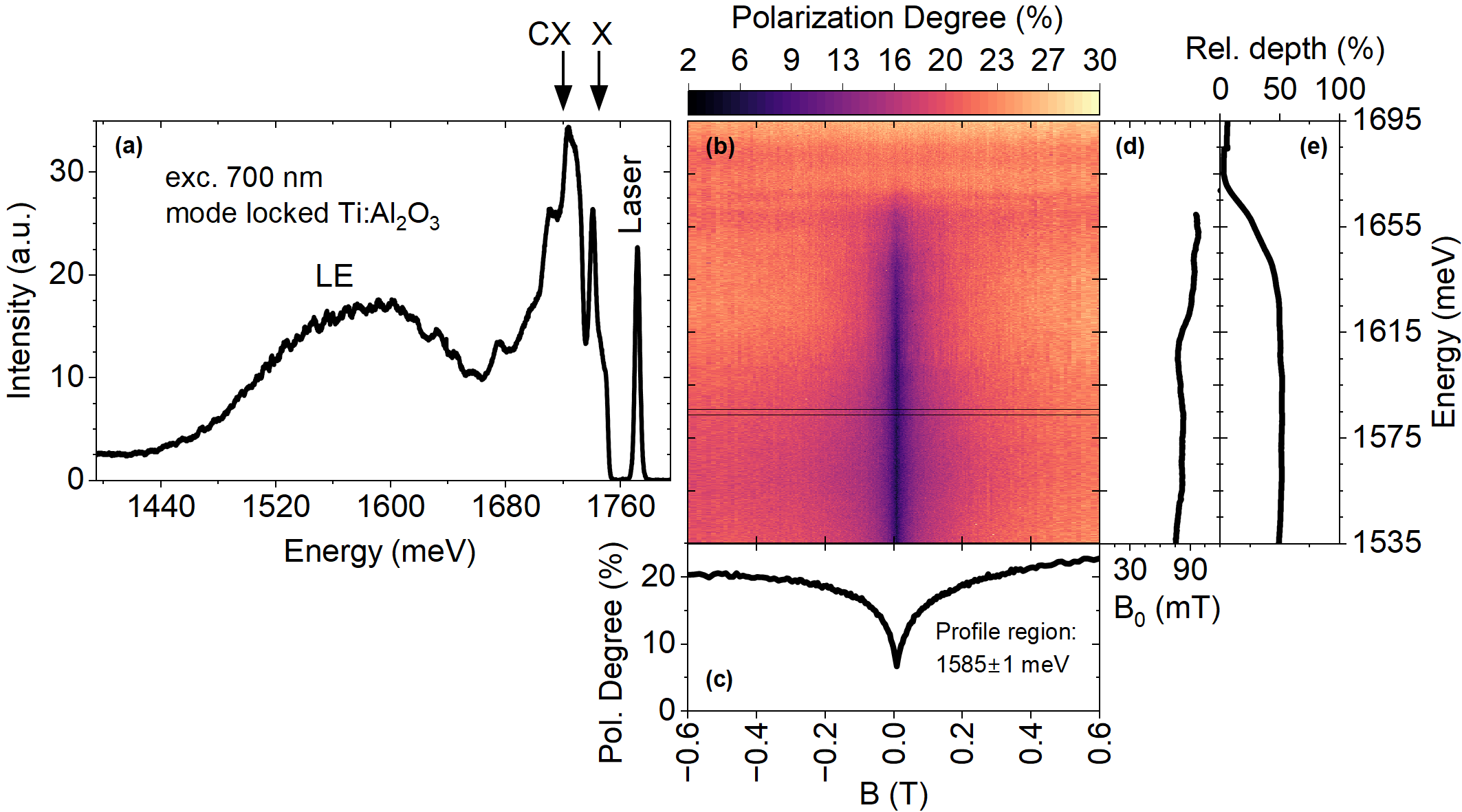}
        \caption{\textbf{(a)} Characteristic emission spectra at T=5 K under 700 nm pulsed excitation, highlighting the Local Exciton (LE) band Charged Exciton (CX) peaks (note the distortion of the CX peak due to the long-pass filter's spectral proximity). \textbf{(b)} Map of the polarization degree, with the profiling region delimited by solid black lines. \textbf{(c)} Mean polarization degree calculated from the marked region. \textbf{(d)} Plot of the $B_0$ (HWHM) of the observed dip versus detected photon energy, demonstrating a stable value of $B_0$ that indicates the FIPE effect is independent of the spectral region within the LE band. \textbf{(e)} Relative depth of the dip as a function of photon energy, showing a clear decrease to zero at the energy corresponding to the cessation of the FIPE effect.}
        \label{map}
    \end{figure*}
    
    The first step of our investigation was to establish the presence of the FIPE effect for the studied MBE-grown samples. During the experiment, two spectra were recorded for each magnetic field value: one with the detection polarization matching the excitation polarization (co-circular, $I_\mathrm{co}$), and one with the opposite polarization (cross-circular, $I_\mathrm{cross}$). 
    
    All the recorded spectra featured excitonic emission band, characteristic for \ce{WSe2} monolayers \cite{oldest}, as seen in Fig. 1a. In the spectrum we distinguish the emission related to the recombination of the neutral exciton (X) at $E = 1.750$ eV, the charged exciton (CX) at $E = 1.717$ eV and the wide band denoted as LE in the range of $E = 1.450$ to $1.650$ eV. The former band is characteristic for tungsten-based dichalcogenides (\ce{WSe2}, \ce{WS2}) and is related to the sign spin-orbit splitting of the conduction band which renders the optical recombination of the lowest energy excitons spin-forbidden. The LE bands consists of both phonon-assisted recombination of the dark exciton, which are particularly prominent in hBN-encapsulated monolayers \cite{encapsulated-peak}, as of the excitons localized on various defects \cite{LE-on-defects}. More comprehensive discussion of the optical properties of these samples, including its temperature dependence, is presented in \cite{Julka}.

    In order to express the difference between the two measured polarization configurations (co- and cross-), for each pairs of spectra we calculate the degree of circular polarization (PD) using the standard formula:
      \begin{equation}
        PD = \frac{I_\mathrm{co}-I_\mathrm{cross}}{I_\mathrm{co}+I_\mathrm{cross}}  \label{eq:hanle}
      \end{equation}
    A representative dependence of this polarization degree as a function of photon energy and the magnetic field is shown in Fig. 1(b-c). The data clearly show the characteristic dip centered at zero magnetic field, which illustrates that application of the magnetic field, regardless of its sign, increases the degree of the PL polarization (FIPE effect).
    
    To quantitatively compare our results with the earlier studies of FIPE \cite{smolen1}, we used the same parametrization of the field dependence of the polarization degree in form of a lorentzian:
    \begin{equation}
    PD = y_0 \cdot \left(1 - \frac{A_\mathrm{mod}}{1+ \left(B-B_\mathrm{rem}\right)^2/B_0^2}\right)
    \end{equation}
    where $y_0$ parameter corresponds to an asymptotic polarization degree value, $A_{mod}$ is a percentage relative amplitude of the dip, $B$ stands for magnetic field induction, $B_0$ corresponds to a \textit{HWHM} of the dip and $B_\mathrm{rem}$ representing remanent field of superconducting coil magnet. We note that the experimental data does not exactly correspond the lorentzian lineshape, but we continue to use it for the sake of compatibility with the earlier research \cite{smolen1}. 

    Figures 1(d) and 1(e) present a typical dependence of the extracted width ($B_0$) and the amplitude of the effect ($A_\mathrm{mod}$). This data exhibits the qualitative features as the FIPE previously reported for exfoliated WSe$_2$ and WS$_2$. In particular, the effect is pertinent to the spectral region of the localized excitons --- the degree of polarization strongly increases with the magnetic field value for all the LEs and is practically constant in the range of X and CX emission. Importantly, all the LEs share almost the same value of characteristic magnetic field $B_0$, which was foundational for identification that FIPE effect reveals the polarization (valley) interplay at the relatively long-lived intermediate state, which precedes the various LEs states in the relaxation cascade \cite{smolen1}. As these characteristics of the observed effect are the same for presently studied samples, we conclude that the FIPE effect here has the same origin. \emph{I.e.}, there exists a stage of the relaxation cascade, which at $B=0$ T effectively quenches the valley polarization, \emph{e.g}., due to intervalley scattering of the excitons. With application of the external magnetic field $B$ we introduce a detuning between the two valleys and this scattering becomes less effective, leading to the excitons retaining more valley polarization.

    \begin{figure}[htb]
        \centering
        \includegraphics[width=0.9\linewidth]{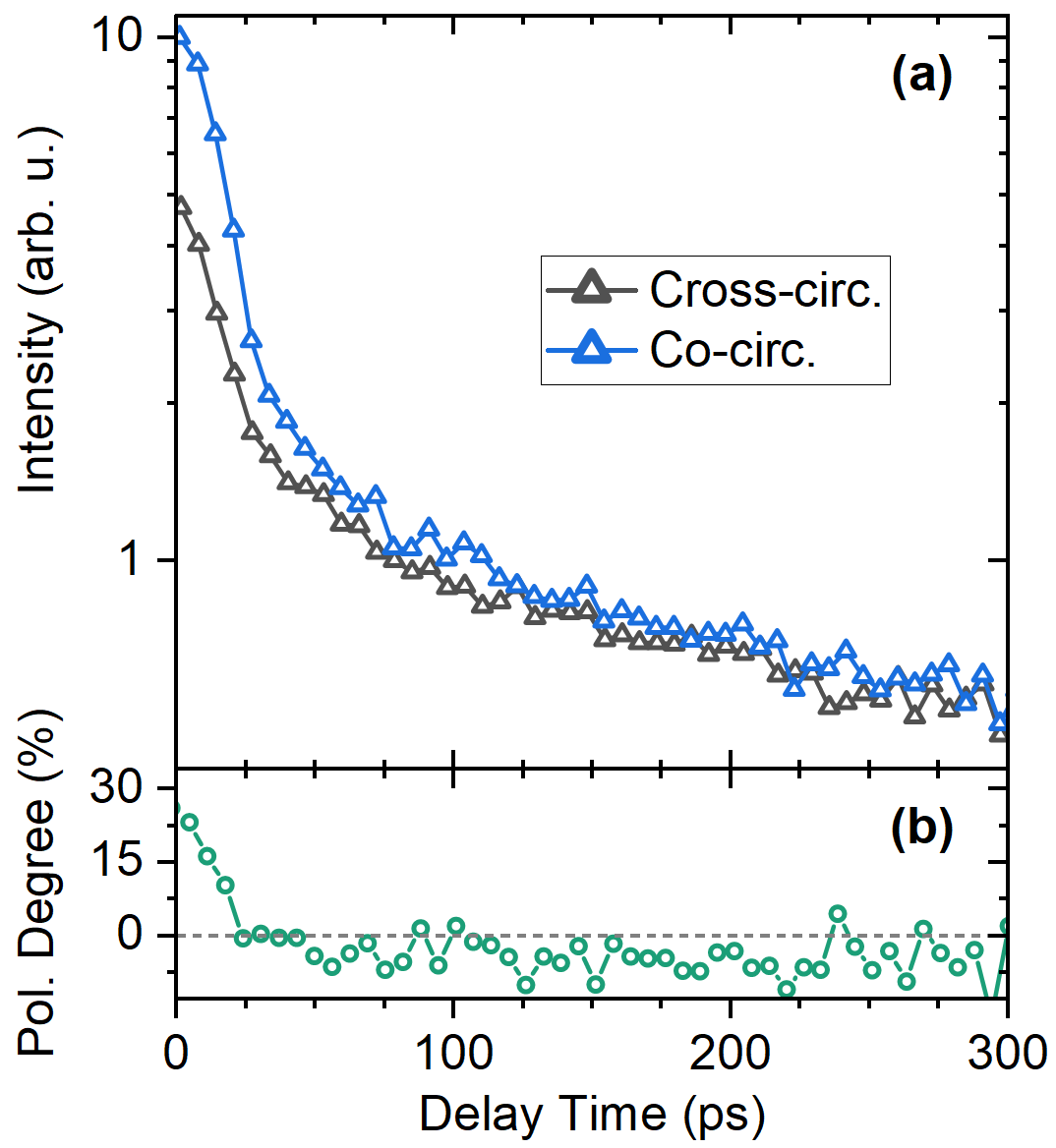}
        \caption{\textbf{(a)} Time-resolved photoluminescence (PL), acquired using a streak camera at $B = 0 \, T$ in the two circular polarization detection configurations (co- and cross-polarized). The panel \textbf{(b)} shows the time-resolved polarization degree.}
        \label{streak}
    \end{figure} 

    Apart from qualitative similarities, our analysis  of the fitted parameter $B_0$ reveals a significant quantitative difference. The value of $B_0$ obtained for the MBE-grown samples is approximately 80 mT at 5 K, which is 4x larger than the $B_0$ value of approximately 20 mT reported for mechanically exfoliated samples \cite{smolen1}. This quantitative difference implies a major change in the valley dynamics. According to the Hanle-like model behind the equation \ref{eq:hanle}, the half-width $B_0$ is directly proportional to the pseudospin depolarization rate $\gamma_\mathrm{depol} = 1 / \tau_{depol}$ at the intermediate stage (in which depolarization occurs). The four-fold increase in $B_0$ therefore indicates the same increase in the depolarization rate and a corresponding four-fold decrease in the depolarization time. Compared to about $\approx$100 ps depolarization time estimated for exfoliated samples \cite{smolen1}, it yields the estimated valley depolarization time for the MBE samples down to approximately 20 ps. 
    
    We note that the observed shortening of the valley depolarization time is in line with general trend of faster dynamics in the MBE-grown samples, as found in other experiments such as  excitation correlation spectroscopy \cite{kacper}.
    
    To complement the CW approach, we have additionally performed time-resolved measurements of the polarization-resolved PL. Similarly to the CW experiment, the PL transient was recorded independently for the $\mathrm{co}-$ and $\mathrm{cross}-$ configurations of the circular polarization. Each of the PL transients shows a multi-exponential decay dynamics, as illustrated in Fig. 2(a) and reported also in other studies \cite{multiexp}. More informative is the transient of the calculated polarization degree, which is presented in Fig. 2(b). The data clearly shows that shortly after the laser excitation, the emitting excitons exhibit higher degree of polarization than at later time. The decay time of this polarization is consistent with the value of 25 ps, earlier inferred based on the CW measurements, corroborating our analysis.

    \begin{figure*}[htb]
            \centering
            \includegraphics[width=0.85\linewidth]{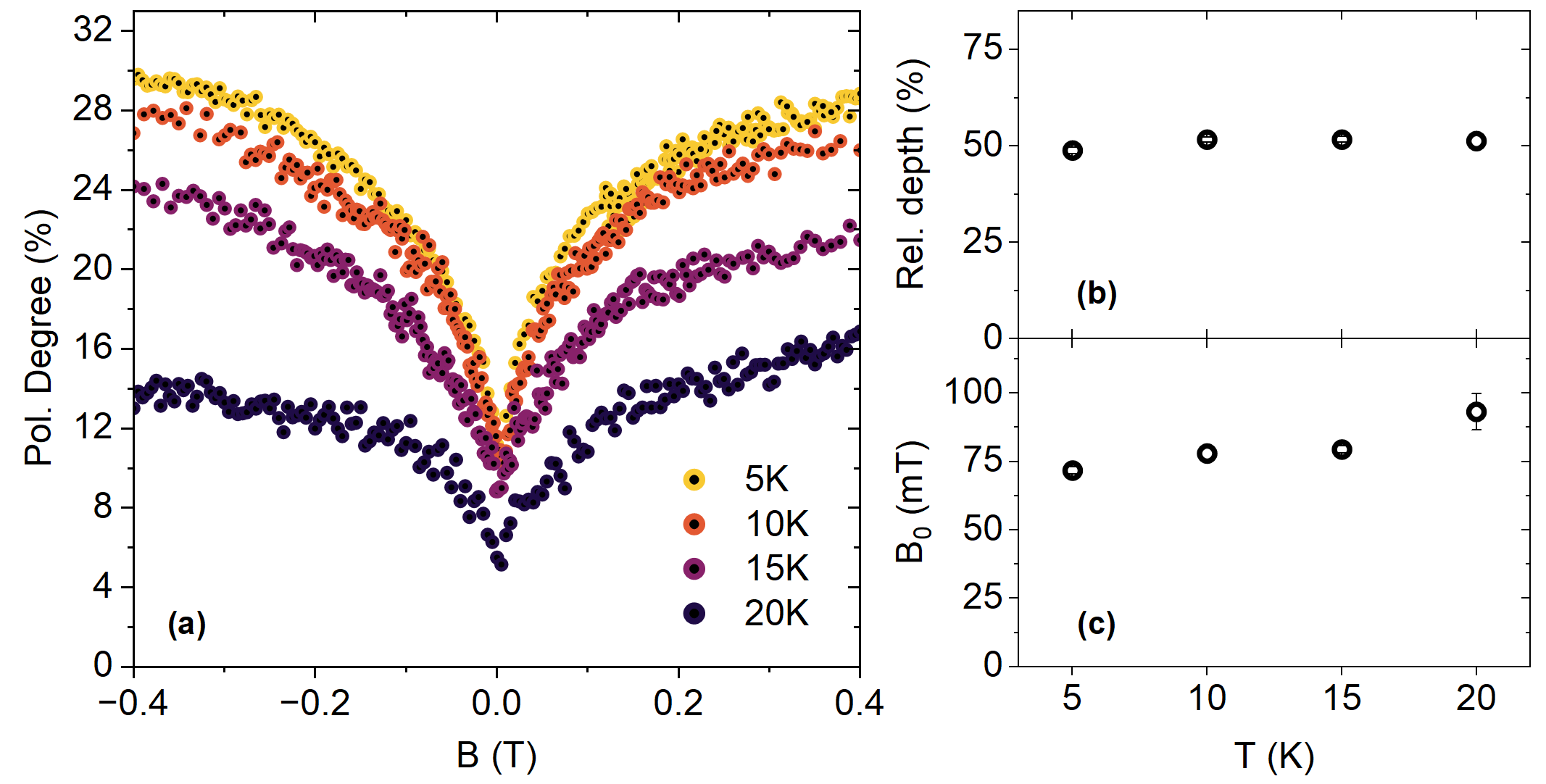}
            \caption{\textbf{(a)} Degree of polarization profiles for different temperatures, averaged over the detection energy range of $1570 \pm 3 \, \unit{meV}$. \textbf{(b)} Relative depth and \textbf{(c)} $B_0$ (HWHM) extracted from the profiles shown in (a).}
            \label{temp}
    \end{figure*} 
    
    As the final experiment, we analyzed how the observed FIPE effect is affected upon elevating the temperature. This experimental series was performed again in CW regime. The representative data obtained for temperatures between 5 and 20 K are presented in Fig. 3(a). The primary effect of the increasing temperature, apart from overall quenching of the PL intensity, is the reduction of the polarization degree. It is intuitive that elevated temperatures accelerates the inter-valley relaxation rate leading to such an effect. Less expected was the fact that the relative amplitude of the FIPE effect ($A_\mathrm{mod}$) is constant in this temperature range (see Fig. 3(b)), while the mechanically exfoliated WSe$_2$ samples already exhibited a visible amplitude drop in this temperature range \cite{smolen2}. Particularly interesting is the width $B_0$ presented in Figure 3(c), which is also practically temperature-independent. Compared with the earlier findings \cite{smolen2}, such a behavior was characteristic for WS$_2$ samples rather than WSe$_2$ samples. More specifically, the observed lack of change in $B_0$ upon increasing the temperature may be explained by assuming that the presently studied samples correspond to the regime when the energy relaxation rate $\gamma_\mathrm{relax}$ is still much faster than the intrinsic inter-valley scattering $\gamma_\mathrm{inter}$, as discussed in Ref. \cite{smolen2}. Again, such an accelerated relaxation dynamics could be expected due to more disordered structure of the MBE-grown flakes \cite{MBE-character, MBE-accel}.

\section{Conclusion}
    This study confirms the presence of magnetic field-induced valley polarization enhancement in high-quality \ce{WSe2} monolayers grown by molecular-beam epitaxy on hBN substrates. While the qualitative effect is similar to that observed in mechanically exfoliated samples, the quantitative analysis reveals a crucial difference: the pseudospin depolarization time in the MBE samples is approximately 20 ps, which is four times shorter than the values reported for exfoliated materials. This finding is supported by time-resolved measurements that show a rapid initial decay of the degree of polarization. The results demonstrate that while advanced epitaxial growth methods can significantly improve optical quality and uniformity, they introduce a distinct defect landscape that fundamentally alters the dynamics of valley relaxation. Therefore, a comprehensive understanding of the interplay between the material's synthesis method and its intrinsic properties is essential for the future design and optimization of valleytronic devices.

\section{CRediT authorship contribution statement}
\noindent
\textbf{Maksymilian Kuna} -- Writing - original draft, Investigation, Formal analysis \\
\textbf{Mateusz Raczyński} -- Writing - review \& editing, Investigation, Supervision, Validation \\
\textbf{Julia Kucharek} -- Preparation of samples \\
\textbf{Takashi Taniguchi} -- Resources \\
\textbf{Kenji Watanabe} -- Resources \\
\textbf{Tomasz Kazimierczuk} -- Writing - original draft and review \& editing, Supervision, Validation \\
\textbf{Wojciech Pacuski} -- Preparation of samples \\
\textbf{Piotr Kossacki}- - Writing - review \& editing, Conceptualization, Supervision, Validation

\section{Declaration of competing interest}
The authors declare that they have no known competing financial interests or personal relationships that could have appeared to influence the work reported in this paper.

\section{Acknowledgements}
We acknowledge the financial support from National Science Centre Poland projects no. 2023/49/N/ST11/04125, no. 2021/41/B/ST3/04183, no. 2020/39/B/ST3/03251 and no. 2020/38/E/ST3/00364. 

K.W. and T.T. acknowledge support from the JSPS KAKENHI (grant numbers 21H05233 and 23H02052) and the World Premier International Research Center Initiative (WPI), MEXT, Japan.

\bibliographystyle{elsarticle-num}
\bibliography{biblio}

\end{document}